\documentclass[aps,prb,twocolumn,groupedaddress]{revtex4-1}
\usepackage{graphicx}
\usepackage{float}
\usepackage{amsmath}
\usepackage{multirow}

\begin{document}


\title{Hyperspatial optimisation of structures}

\author{Chris J.\ Pickard}
\email[]{cjp20@cam.ac.uk}
\affiliation{Department of Materials Science \& Metallurgy, University of Cambridge, 27 Charles Babbage Road, Cambridge CB3~0FS, United Kingdom}
\affiliation{Advanced Institute for Materials Research, Tohoku University 2-1-1 Katahira, Aoba, Sendai, 980-8577, Japan}


\date{\today}

\begin{abstract}
Anticipating the low energy arrangements of atoms in space is an indispensable scientific task. Modern stochastic approaches to searching for these configurations depend on the optimisation of structures to nearby local minima in the energy landscape. In many cases these local minima are relatively high in energy, and inspection reveals that they are trapped, tangled, or otherwise frustrated in their descent to a lower energy configuration. Strategies have been developed which attempt to overcome these traps, such as classical and quantum annealing, basin/minima hopping, evolutionary algorithms and swarm based methods. Random structure search makes no attempt to avoid the local minima, and benefits from a broad and uncorrelated sampling of configuration space. It has been particularly successful in the first principles prediction of unexpected new phases of dense matter. Here it is demonstrated that by starting the structural optimisations in a higher dimensional space, or hyperspace, many of the traps can be avoided, and that the probability of reaching low energy configurations is much enhanced. Excursions into the extra dimensions are progressively eliminated through the application of a growing energetic penalty. This approach is tested on hard cases for random search -- clusters, compounds, and covalently bonded networks. The improvements observed are most dramatic for the most difficult ones. Random structure search is shown to be typically accelerated by two orders of magnitude, and more for particularly challenging systems. This increase in performance is expected to benefit all approaches to structure prediction that rely on the local optimisation of stochastically generated structures.
\end{abstract}


\maketitle

\section{Introduction}

The possible existence of more, or indeed fewer, spatial dimensions than the usual three inspire theoretical physics,\cite{randall2002extra} and literature.\cite{flatland,threebody} String theories postulate large numbers of additional dimensions (wrapped up on themselves in such a way that we are oblivious to them). In condensed matter physics, the interaction of particles in infinite dimensions can give a good approximation to their behaviour in three, and be analytically simpler.\cite{metzner1989correlated} Superspaces are used to describe the crystallography of modulated \cite{de1974pseudo} and quasicrystals.\cite{janssen1986crystallography} In biology, it has been proposed that evolution might take hyperdimensional shortcuts.\cite{conrad1990geometry,conrad1992mv,cariani2002extradimensional} Whether or not these extra dimensions are real, they, and their shortcuts, might be created and exploited computationally.\cite{chikenji1999multi,rodinger2005absolute}

The structure of matter at the atomic scale determines its physical properties. Carbon, arranged in the diamond lattice, is extremely hard, and transparent. In layers, as graphite, it is soft and opaque. The prediction of the likely structures that collections of atoms adopt depends on identifying the low energy arrangements of those atoms. \cite{hoff1898arrangement,wales2003energy}  The diamond and graphite structures are among the lowest in energy for carbon, with graphite the lowest, the ground state, and diamond slightly higher, and metastable under normal conditions. A global search for all the low lying, and not just the lowest, minima is needed. This is very challenging, but there has been considerable progress in the first principles (through density functional theory - DFT \cite{car1985unified,payne1992iterative,hasnip2014density}) prediction of material and chemical structure.\cite{pickard2006high,oganov2006crystal,wang2012calypso} These advances have been made possible by the development of robust, reliable,\cite{lejaeghere2016reproducibility} and efficient computer codes,\cite{clark2005first,kresse1996efficient} and the rapid growth and availability of computational resources. 

For even small systems, there can be a large number of local minima which are relatively high in energy, and frustrate the search. This can be easily seen to be the case for strongly covalently bonded systems, such as carbon. Once the covalent bonds have formed, a large barrier to their rearrangement is created. Even if lower energy configurations are nearby, they cannot be reached. The system is trapped in this potentially highly energetic conformation.

In addition to classical and quantum annealing,\cite{schon2001determination,santoro2002theory}  evolutionary,\cite{deaven1995molecular,oganov2006crystal} swarm\cite{wang2012calypso} or basin/minima hopping\cite{wales1999global,goedecker2004minima} based algorithms have been employed in an attempt to escape the local minima. They learn from prior explorations of configuration space, and focus their computational effort on the probed low energy regions. The structures generated at each step are optimised to nearby local energy minima, using gradient based algorithms. Learning algorithms can be intricate, and require the careful choice of parameters. A simple alternative is random structure search (RSS, and from first principles, Ab Initio Random Structure Searching, or AIRSS), the repeated local optimisation of stochastically generated structures.\cite{pickard2006high,pickard2011ab}

In this manuscript a modification to local geometry optimisation is presented in which extra spatial dimensions are created, and subsequently eliminated. The probability of being trapped in high energy local minima is shown to be greatly reduced. This is achieved without any special preparation of the initial structures, and is relatively insensitive to the few parameters that are required. The performance of the geometry optimisation of structures from hyperspace (or GOSH) is demonstrated through its application to structure search on model energy landscapes. The prospects of performing these accelerated searches at the level of accuracy provided by first principles methods are discussed.

\section{Random structure search}

AIRSS is a straightforward, and effective, approach to first principles structure prediction.\cite{pickard2006high,pickard2011ab} It is sampling based,\cite{white1971survey} intrinsically parallel, and well adapted to modern computer architectures. The random, or stochastic, generation of structures ensures uncorrelated results. There is no attempt to avoid traps, but the coverage of the accessible region of configuration space is broad. It has proven to be particularly well suited to the prediction of unexpected phases of dense matter (a regime in which our chemical understanding is still developing). Mixed molecular and atomic-like phases of hydrogen identified using AIRSS\cite{pickard2007structure} provided an excellent model for the then unknown phase IV of hydrogen.\cite{howie2012mixed} Aluminium was found to adopt surprisingly complex incommensurate structures, \cite{pickard2010aluminium} and ammonia to form ionic ammonium amide.\cite{pickard2008highly,palasyuk2014ammonia,ninet2014experimental} The possibly surprising success of AIRSS is likely due to its exploitation of natural features of the first principles energy landscape itself, in particular the landscape's relative smoothness, and its arrangement as a fractal packing of basins, with a power law distribution in volumes.\cite{massen2007power,pickard2011ab}

\begin{figure}[]
\centering
\includegraphics[width=0.45\textwidth]{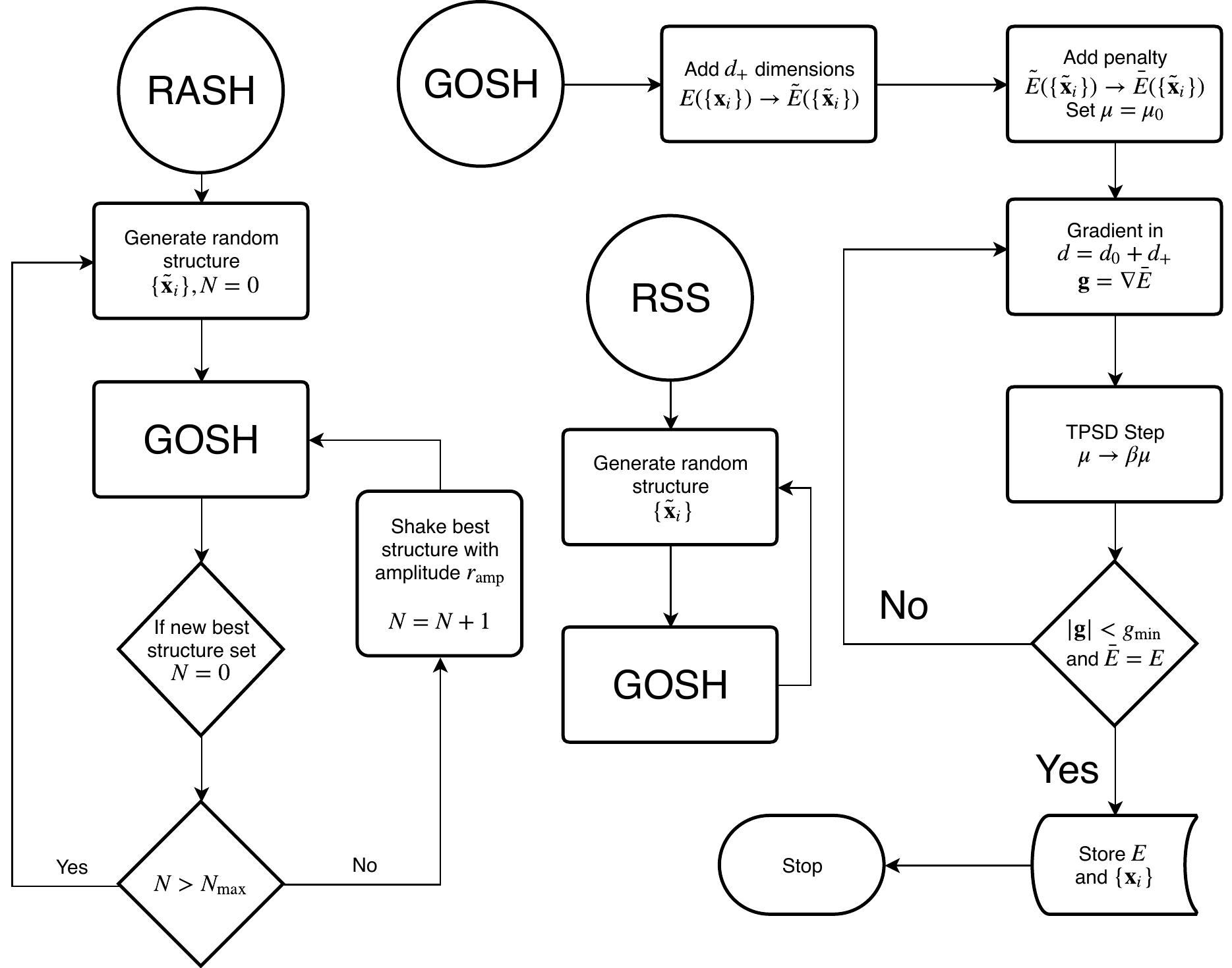}
\caption{Flowcharts for the Geometry Optimisation of Structures from Hyperspace (GOSH), Random Structure Search (RSS) using GOSH, and Relax and Shake (RASH) using GOSH.}
\label{flow}
\end{figure}
There are situations in which the probability of encountering low energy configurations in a random search is relatively small, and there are a large number of traps. These include ensuring well optimised arrangements for both the core and surface atoms in clusters, finding the correct ordering in multispecies systems, and covalent networks of strongly directional bonds such as might be found in diamond and complex biological or organic molecules. These challenges can be reduced by preparing the initial random structures appropriately, restricting the search to the regions of configuration space that are presumed to contain the lowest energy structures.\cite{pickard2011ab}  Atoms can be confined in spheres, or ellipsoids, to encourage well packed, and presumably low energy, clusters. Species dependent minimum separations, fragments and coordination constraints can be used to generate structures that are compatible with the known chemistry of the system.\cite{pickard2011ab,shi2018stochastic} In addition, the fact that low energy configurations typically exhibit some symmetry can be exploited through the imposition randomly chosen symmetry operations on the structures. Randomly selecting from these ``sensible'' structures enables the successful application of AIRSS to genuinely difficult and realistic systems, such as grain boundaries and interfaces, \cite{schusteritsch2014predicting} or complex carbon structures.\cite{shi2018stochastic}  In what follows we explore an approach that does not require such careful preparation of initial structures for challenging systems.

\section{Extra spatial dimensions}

Structure prediction involves the exploration of highly dimensioned \emph{configuration spaces}, with total dimensionality of $Nd$, where $N$ is the number of atoms and $d$ is the number of \emph{spatial dimensions} that those atoms inhabit. A distinction is to be drawn between these configuration spaces, and \emph{hyperspaces} with additional spatial dimensions for the atoms to move about in. In what follows the number of dimensions, $d$, of a hyperspace is given as $d=d_0+d_+$, where $d_0$ is the dimensionality of the \emph{normal} space, and $d_+$ is the number of extra spatial dimensions. If the normal space is two dimensional ($d_0=2$, a plane) and $d_+=1$, then $d=2+1=3$ and the resulting hyperspace is three dimensional.  For a three dimensional normal space, a hyperspace would have four or more dimensions.

\section{Optimisation from Hyperspace}

\begin{figure}[]
\centering
\includegraphics[width=0.45\textwidth]{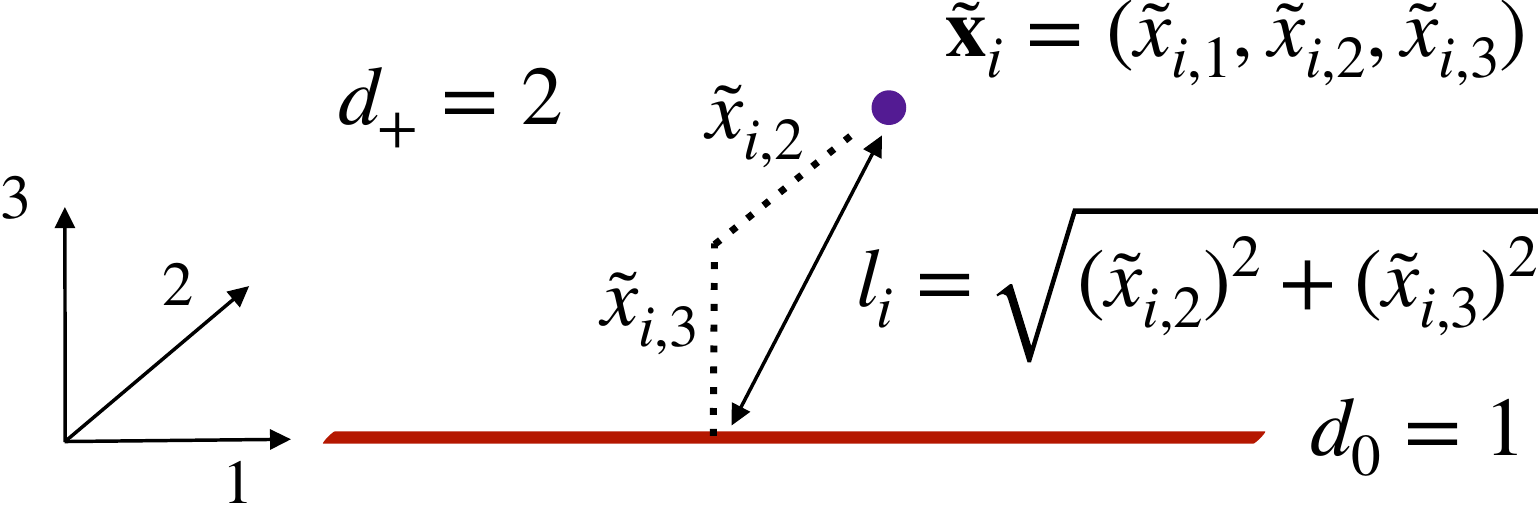}
\caption{Defining the distance, $l_i$, of atom $i$ in hyperspace to normal space, for $d=d_0+d_+=1+2$.}
\label{distance}
\end{figure}

GOSH is a scheme for exploiting hyperspace to avoid traps in the energy landscape during structural optimisation and is described as a flowchart in Figure \ref{flow}. The energy landscape in the normal space, $E(\{\bf{x}_i\})$, is first extended to hyperspace, $\tilde E(\{\bf{\tilde x}_i\})$, where $\{\bf{x}_i\}$ and $\{\bf{\tilde x}_i\}$ are the positions of the atoms in the normal and hyperspaces respectively. A structural optimisation is then initiated from a structure stochastically generated in hyperspace. As it progresses an increasing energetic penalty  is applied, with a strength related to the distance of the atoms in the structure from the normal space. The distance of atom $i$ from the normal space, $l_i$, is computed (see Figure \ref{distance}) as: 
\begin{equation}
l_i=\sqrt{\sum_{\delta=d_0+1}^{d}(\tilde x_{i,\delta})^2}.
\label{defli}
\end{equation}
The extension of the energy to hyperspace is not uniquely defined. The extended energy landscape should be identical to the normal energy landscape if the structure is entirely in the normal space (i.e. the distances of all the atoms from normal space are zero). The extension should also, in some sense, be physically reasonable. There is a large class of interactions that depend only on the distances between atoms, including the two-body Coulomb and Lennard-Jones potentials. A natural scheme for the extension of the energy landscape to hyperspace is to simply replace these distances in normal space by those calculated in hyperspace. Although not developed further here, three-or-more body angle dependent interactions might similarly be extended by evaluating those angles in hyperspace.

The atoms should be free to explore the hyperspace at the early stages of the optimisation. But, as the final configuration must exist entirely in normal space, some computational means of enforcing this is required. One approach would be to introduce a hard wall potential which moves towards zero from above and below in the extra dimensions and compresses the structure into normal space. Instead, the approach used here is to add a harmonic term to the total energy of the system, which depends on the distance the atoms have entered into the extra spatial dimensions. As the strength of this penalty is increased it becomes more difficult for the atoms to move into the extra dimensions, and they are progressively confined to the normal space. The penalised energy extended to hyperspace is:
\begin{eqnarray}
\bar E(\{{\bf \tilde x}_i\})&=&\tilde E(\{{\bf \tilde x}_i\})+\frac{1}{2}\mu\sum_{i}l_i^2 \nonumber\\
&=&\sum_{i,j}V_{ij}(\tilde r_{ij})+\frac{1}{2}\mu\sum_{i}l_i^2,
\end{eqnarray}
where $\tilde r_{ij}=|\tilde {\bf x}_i-\tilde {\bf x}_j|$ is the distance between atom $i$ and atom $j$ in the hyperspace and $V_{ij}(\tilde r_{ij})$ is the pair interaction potential used in the examples following. The gradients of $\bar E(\{{\bf \tilde x}_i\})$ with respect to $\{{\bf \tilde x}_i\}$ are those used for the structural optimisation.

\begin{figure}[]
\centering
\includegraphics[width=0.45\textwidth]{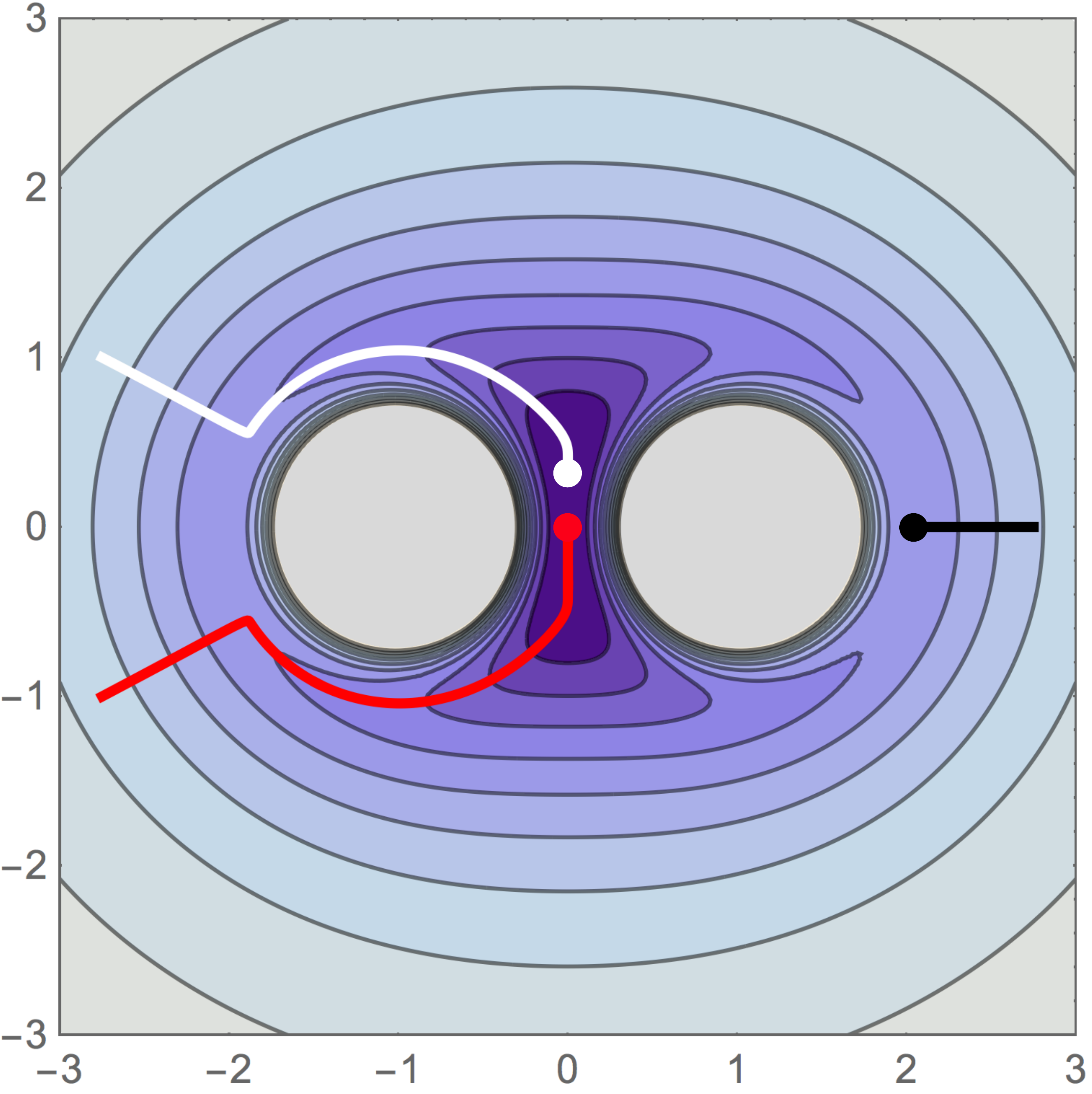}
\caption{A Lennard-Jones trimer for $d=1+1$. Two identical  Lennard-Jones 4-2 atoms (with $\sigma=1$) are fixed at $(\pm1,0)$, and a third atom seeks its lowest energy. Steepest descent paths from i) $(2.75,0)$, constrained to $d_0$ (black), ii) $(-2.75,1)$ in $d=1+1$ (white), and iii) $(-2.75,-1)$ in $d=1+1$ with increasing $\mu$ (red).}
\label{gosh}
\end{figure}

A barrier that is unavoidable in one dimension may be circumvented in two or more dimensions.  This observation motivates the current scheme. In Figure \ref{gosh} a toy system is described which demonstrates the potential of taking excursions into hyperspace to avoid traps in the energy landscape. The normal space is one dimensional. The global minima in this one dimension is between the two fixed atoms which are located at $\pm1$. Any optimisation that is started from above $1$ or below $-1$ will become trapped in a metastable state (the black curve). Extending the energy landscape to two dimensions completely changes the situation. The trap becomes a saddle point, and as long as the atom is not precisely confined to the first dimension, one of the degenerate global minima will be found (the white curve). These minima live in the hyperspace but the increasing energetic penalty eventually pushes the atom to the global minimum in the normal space (red curve). The penalty must be increased from a small value, and slowly enough so that the atom is not immediately forced into normal space, and the metastable trap.

In a two dimensional normal space, with one extra dimension ($d=2+1$), at the start of the optimisation the atoms move freely in the full three dimensional hyperspace. As the optimisation proceeds the increasing harmonic penalty leads to a force pushing the atoms into the plane of the normal space, flattening the ensemble. If the atoms are free to do so, they rapidly relax into the plane as the distances of the atoms from the normal space decrease. However, if the atoms push up against each other, leading to significant forces in the third dimension, this collapse is delayed, and the additional dimension continues to be explored. Traps that might exist in two dimensions can still be avoided and the extra freedom is focussed on the parts of the structure that need it.
\begin{figure}[]
\centering
\includegraphics[width=0.45\textwidth]{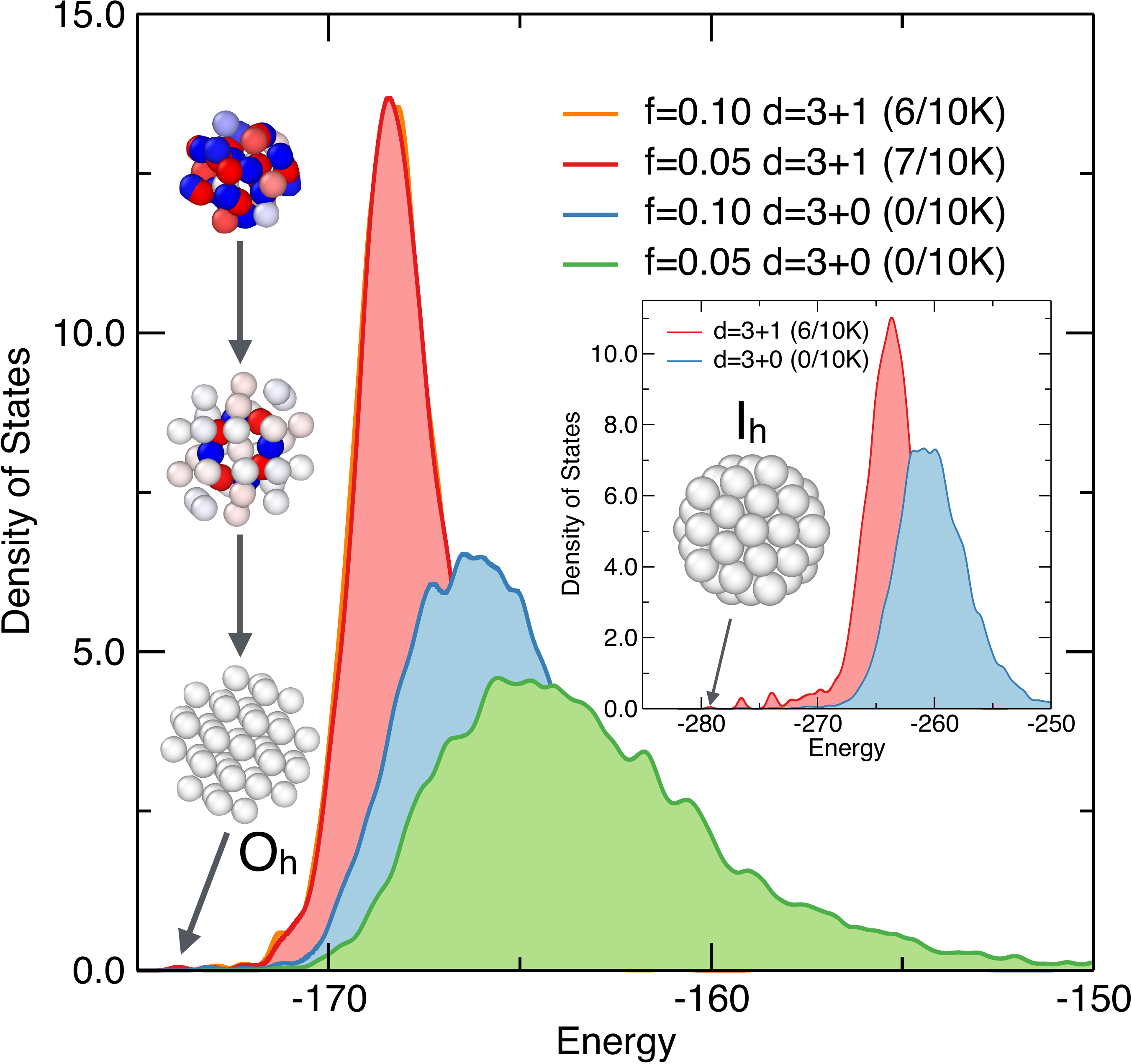}
\caption{Density of structural states for 38 atom Lennard-Jones clusters (inset: 55 atoms, $f=0.1$). Snapshots of an optimisation of a random initial structure with $d_{+}=1$, whose local minima corresponds to the $O_h$ non-icosahedral ground state are presented. The atoms are colour coded according to their location in the extra fourth dimension: increasingly red for positive values, and blue for negative. The probability of encountering the ground state is provided in brackets. Note that for high packing densities with $d_+=0$ the $O_h$ ground state is generated with high probability, and so the peak in the structural density of states is expected to shift further leftwards with increasing $f$.}
\label{lj}
\end{figure}

\section{Implementation and testing}
\label{iat}

To understand the performance of GOSH, computational experiments on an implementation for finite collections of atoms interacting through simple potentials are performed. The initial random structures are generated using the following procedure for all the test cases presented below. Each atom is surrounded by a hard hypersphere, with a radius determined by the interaction potential (for example, half the expected equilibrium bond length). A larger, confining, hypersphere is constructed so that its volume is $1/f$ times greater than the summed total volume of the atom centered hyperspheres, where $f$ is the packing fraction. The atoms are then randomly placed into the larger hypersphere so that their hyperspheres do not overlap with each other, or project outside the larger confining hypersphere. As the packing fraction, $f$, is increased the initial random configuration is more densely packed. The maximum possible packing fraction decreases rapidly with an increasing number of dimensions, which is consistent with what is known about the packing of hyperspheres in highly dimensioned spaces.\cite{skoge2006packing} The random structures are thus generated with a blue noise distribution, and are a form of Poisson disk sampling.\cite{bridson2007fast}

\begin{table}[]
\centering
\caption{The mean negative log-probability of encountering the ground state (GS), $-\log_{10}(p_e)$, for a selection of Lennard-Jones clusters with different sizes. Means based on more than 100 encounters, except for those marked $\dag$: more than 10 encounters. The point group (PG) for the established global minima are indicated. \cite{wales1997global} The initial dimensionality, $d$, is given as $d_{0}+d_{+}$. }
\label{ljclusters-2}
\begin{tabular}{|c|c|c|c|c|c|c|c|}
\hline
\multirow{2}{*}{d}      & \multirow{2}{*}{f} & \multicolumn{6}{c|}{Num. of Lennard-Jones atoms (PG of GS)} \\ \cline{3-8} 
                                 &                             & 37  ($C_1$)  & 38  ($O_h$)  & 39 ($C_{5v}$) & 47 ($C_1$) & 55 ($I_h$) & 69 ($C_{5v}$) \\ \hline
\multirow{3}{*}{3+0} & 0.05                     & 4.5 & 6.0 & 4.4 & 4.6 & 5.6  & 7.0$^\dag$ \\ 
                                 & 0.10                     & 3.8 & 5.0 & 3.8 & 4.1 & 4.8 & 6.5$^\dag$ \\ 
                                 & 0.20                     & 3.2 & 3.9 & 3.1 & 3.5 & 3.6 & 5.9$^\dag$ \\ \hline
\multirow{3}{*}{3+1} & 0.05                     & 3.3 & 3.4 & 3.1 & 3.6 & 3.3 & 5.9$^\dag$\\ 
                                 & 0.10                     & 3.2 & 3.3 & 3.2 & 3.6 & 3.2 & 5.9$^\dag$ \\ 
                                 & 0.20                     & 3.2 & 3.2 & 3.2 & 3.6 & 3.2 & 5.9$^\dag$ \\ \hline
\multirow{1}{*}{3+2} & 0.10                     & 3.5 & 3.8 & 3.2 & 3.7 & 3.3 & 5.8$^\dag$\\ \hline
\end{tabular}
\end{table}

The two point steepest descent (TPSD) algorithm, due to Barzilai and Borwein,\cite{barzilai1988two} is used to optimise the initial configurations and move them to nearby local minima. It is chosen because it is simple to implement, efficient, and requires no line-minimisation. A small positive initial step size is selected, and the absolute value of the computed step is used for subsequent iterations, to ensure progress towards a minimum (as opposed to a more general stationary point). The TPSD algorithm is not monotonic, and backtracking is not implemented. Furthermore, no attempt is made to ensure that the structure remains in the same basin throughout the optimisation process. Tests of GOSH using gradient descent, with a smaller fixed step size, show similar results if the rate at which the penalty grows is increased to account for the order of magnitude larger number of steps required. Momentum gradient descent,\cite{qian1999momentum} which provides a similar performance to TPSD with smoother convergence, exhibits identical results using identical parameters. The performance of GOSH does not appear to be sensitive to the details of the local optimisation scheme used. For simplicity the TPSD is employed in its non-preconditioned form, although much better performance is to be expected from well preconditioned algorithms for geometry optimisation.\cite{pfrommer1997relaxation,packwood2016universal}
\begin{figure}[]
\centering
\includegraphics[width=0.45\textwidth]{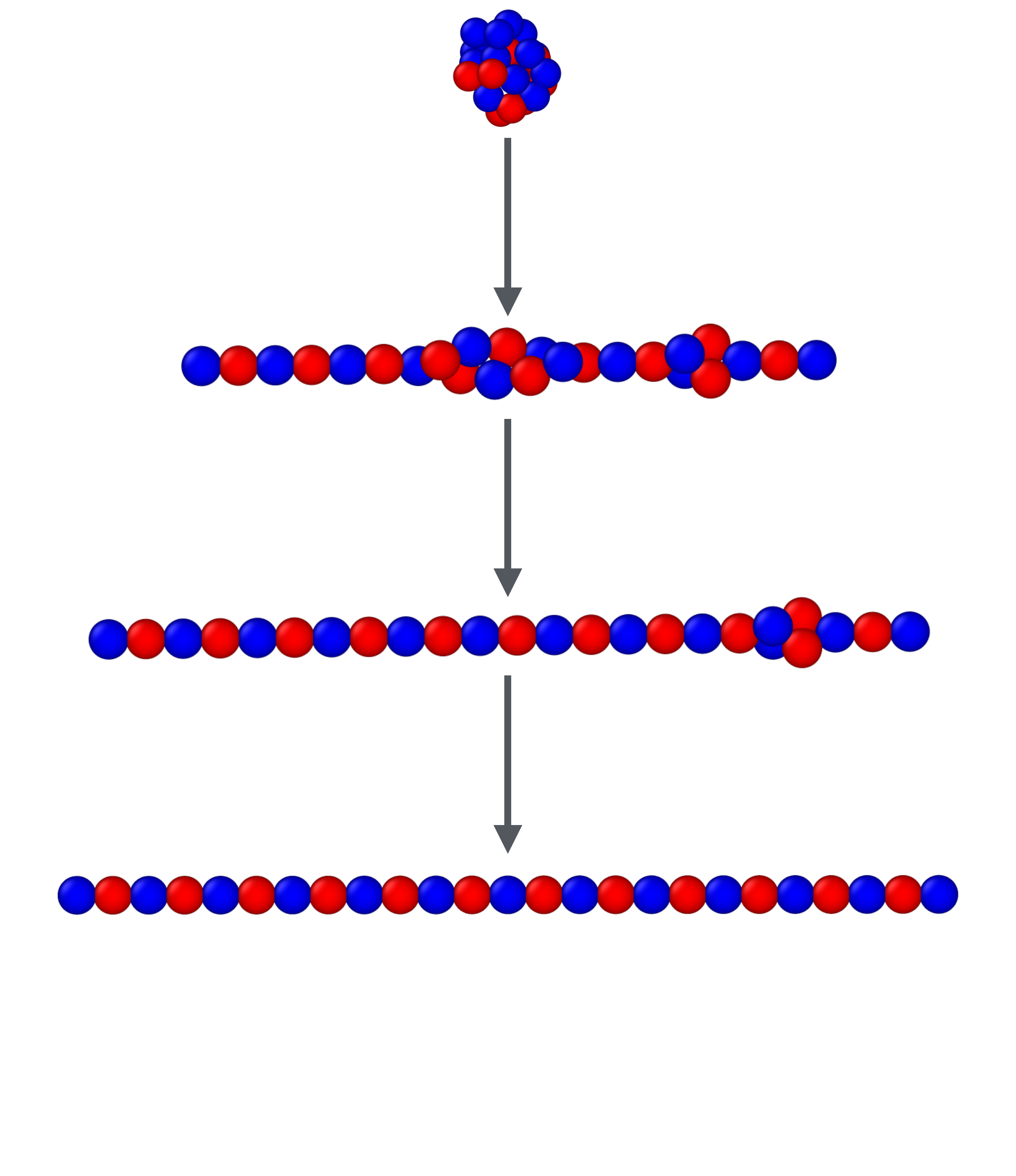}
\caption{The optimisation of an initially random binary structure packed into a three dimensional hypersphere, with $d=1+2$, toward the ground state in $d_0=1$. }\label{wire1}
\end{figure}

Model interatomic potentials are constructed in terms of the distances between pairs of atoms. The distances are defined as the $l^2$-norm in the relavent hyperspace. A Lennard-Jones 12-6 potential ($\epsilon=1$ and $\sigma=1$) is used for the single species cluster tests (Section \ref{ljc}), and a modified form in which like-species interactions are made fully repulsive, by taking the 6-term to be positive, are used for the binary clusters (Section \ref{bs}).\cite{pickard2010aluminium} For the covalently bonded network (Section \ref{cs}), the connectivity is fixed via a predetermined adjacency matrix. The bonded atoms interact through a harmonic potential, with a minimum at the chosen bond-length. The non-bonded interactions are also described by a harmonic potential, with a minimum value at $1.2\sqrt{3}$ times the bond-length, and zero beyond that.

 The strength of the harmonic penalty, $\mu$, is initially given a small value, $\mu_0$. It is increased by a factor of $\beta$ on each cycle of the structural optimisation. The optimisation halts when both the magnitude of the gradient of the energy, $\bar E(\{\tilde {\bf x}_i\})$, is below a threshold, and $\mu$ is greater than some large value, ensuring that the distance of all the atoms from normal space is zero to within an acceptable tolerance and that $\bar E(\{\tilde {\bf x}_i\})=E(\{ {\bf x}_i\})$. The value of $\beta$ should be chosen so that this maximum value of $\mu$ is reached within the typical number of optimisation steps required for convergence. In the following tests, $\beta=1.001$ and $\mu_0=10$, except for the covalently bonded network, where $\mu_0=1$. No particular attempt was made at this stage to tune these values, although for larger systems, which require more optimisation steps, $\beta$ and/or $\mu_0$ should be decreased. It should be noted that as $\mu$ changes with each step of the optimisation there is an inconsistency  between the energy and the gradients, but this has not been found to prevent convergence.
\begin{figure}[]
\centering
\includegraphics[width=0.45\textwidth]{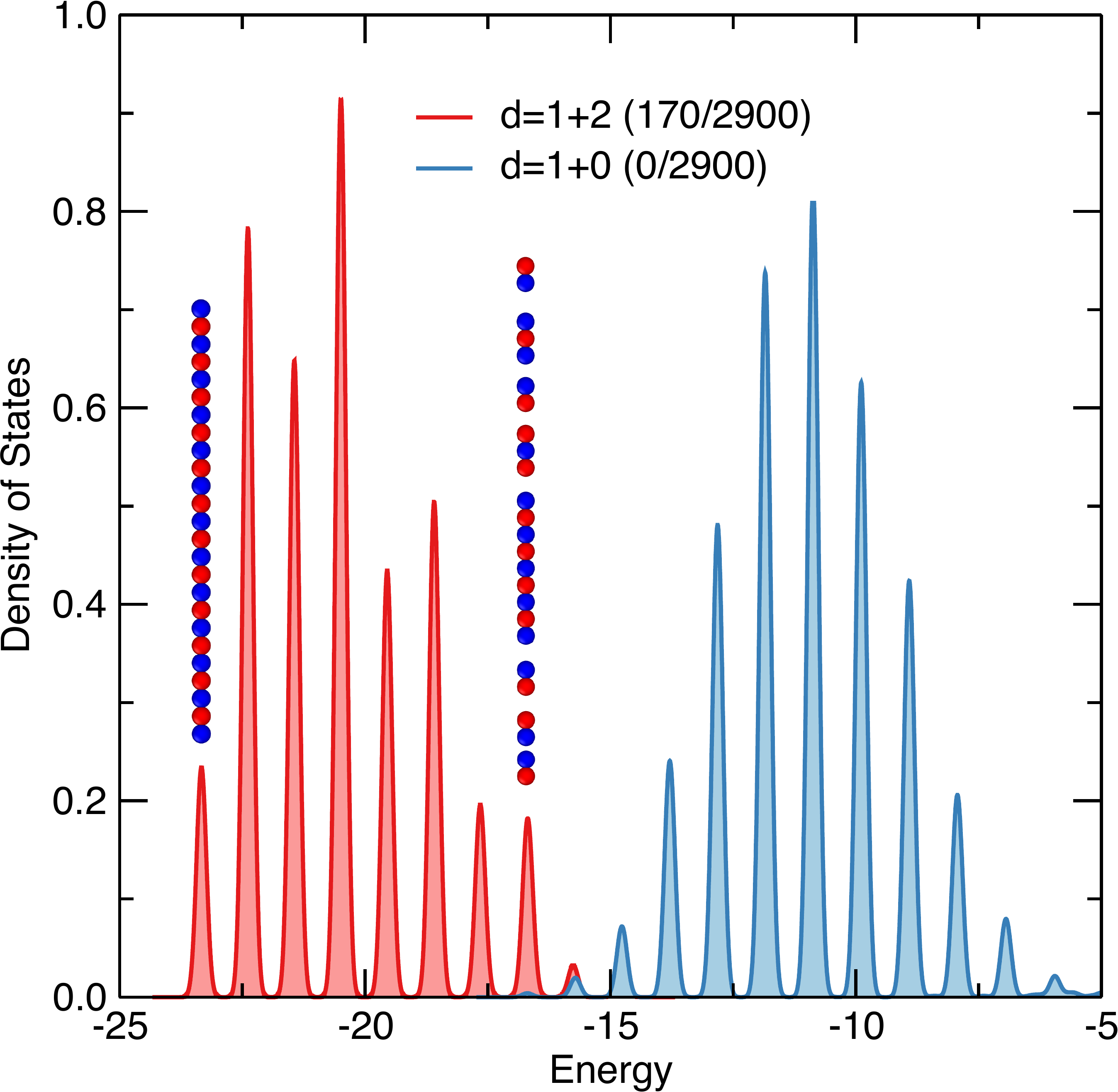}\\
\caption{Density of states for a model 1d binary system, with 12 atoms of type A (red, ``positive''), and 13 atoms of type B (blue, ``negative''). Random structures were generated with $f=0.3$. The gaps between fragments have been reduced for presentational purposes. }\label{wire2}
\end{figure}

\section{Visualisation}

The optimisation process is visualised using the OVITO code,\cite{stukowski2009visualization} which has also been used to generate the figures presented here. For a model 1D chain (see Figures \ref{wire1} and \ref{wire2}), the movement in the additional two dimensions is easy to monitor. But in general it is difficult to visualise objects in hyperspace. The motion of the atoms can be followed for the first three of the dimensions of a given hyperspace using standard 3D visualisation techniques, simply by ignoring the extra dimensions. It is then common to see atoms passing through each other, while they keep their distance in the extra dimensions. This apparent tunneling is exactly the behaviour that  enables the avoidance of traps. In the case of $d=d_{0}+d_{+}=3+1$, the excursions into the fourth dimension can be followed by colouring the atoms according to their component in the extra dimension -- for example, more strongly red as $\tilde x_{i,4}$ becomes more positive, and blue as it becomes more negative. This approach has been taken to generate Figures \ref{lj}, \ref{net} and \ref{net2}.

\section{Results}

\subsection{Lennard-Jones clusters}
\label{ljc}

The low energy structures of small clusters of atoms interacting through the Lennard-Jones potential have been intensively studied, and the global minima have been identified with a high degree of certainty.\cite{wales1997global,leary2000global} As such they present an excellent system on which to test novel methods of optimisation.\cite{schonborn2009performance,pickard2011ab} The performance of the combination of GOSH with RSS has been examined for a range of Lennard-Jones clusters varying in size from 37 to 75 atoms. The 38 and 75 atom clusters are known as difficult cases for structure prediction, exhibiting multiple funnels in the energy landscape.\cite{doye1999double} 

In Figure \ref{lj} the structural density of states for 38 and 55 atom Lennard-Jones clusters are presented. The packing fraction, $f$, strongly influences the density of states for $d_{+}=0$,\cite{jackson2006statistical,locatelli2002fast} but much less so for $d_{+}=1$. 

The evolution of a 38 atom cluster from an initially random structure to the $O_h$ ground state on optimisation with GOSH is shown in Figure \ref{lj}. Initially, most atoms are some distance from the normal space (and so strongly coloured red and blue), but this distance reduces as the optimisation progresses, and the penalty for incursion into the extra and fourth dimension grows. Midway through the optimisation the still frustrated core atoms continue to explore hyperspace, but the outer shell is already to be found entirely in the normal three. The probability of encountering the ground state is much larger for optimisations started in hyperspace for both the 38 and 55 atom clusters.

\begin{figure}[]
\centering
\includegraphics[width=0.45\textwidth]{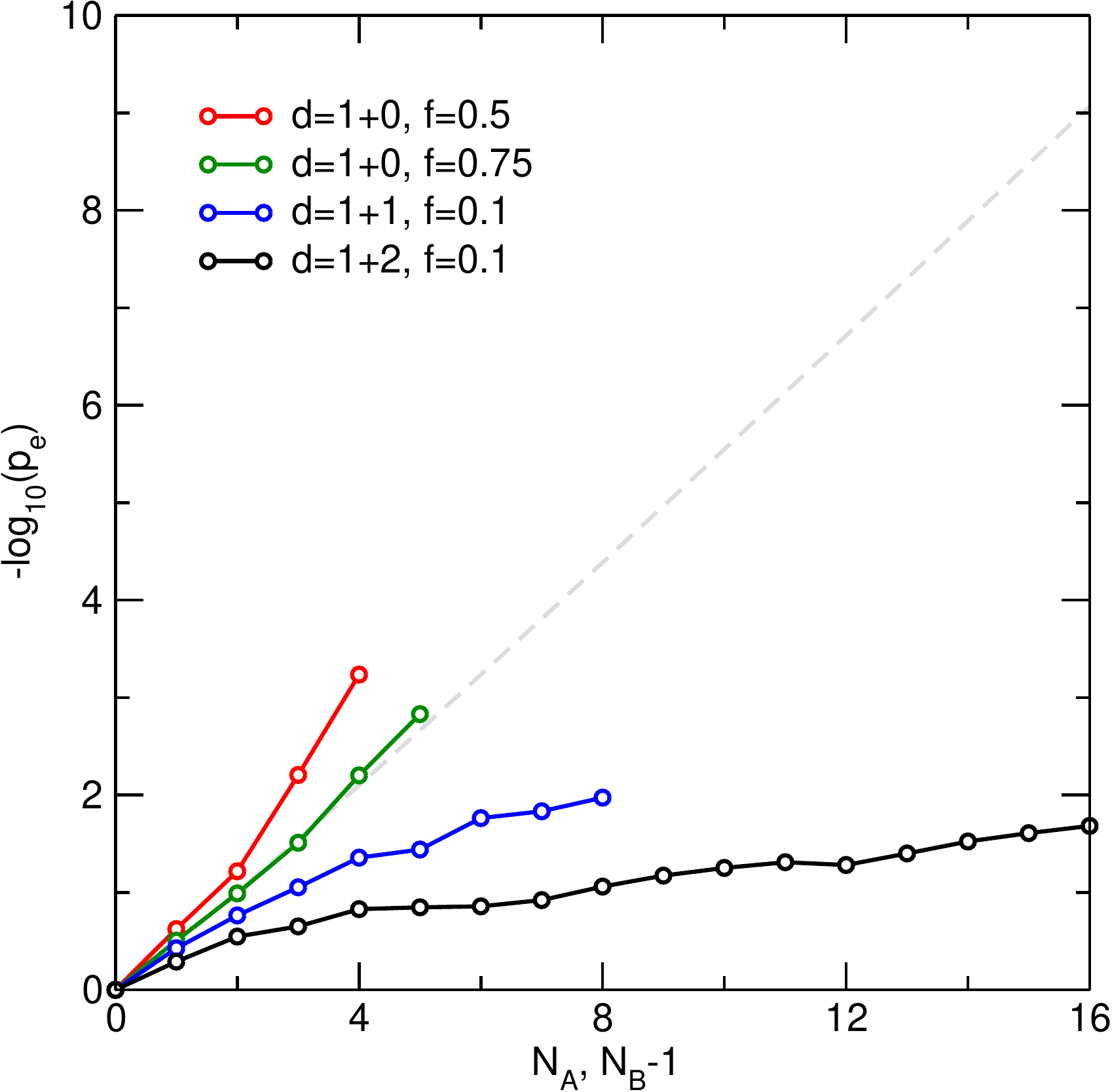}\\
\caption{Evolution of the encounter probability with system size, for the one dimensional ($d_0=1$) binary Lennard-Jones system. The grey dashed line indicates the probablity of randomly choosing the correct ordering.}
\label{wire3}
\end{figure}
In order to generate accurate estimates for the probabilities of encounter, $p_e$, presented in Table \ref{ljclusters-2}, up to $10^8$ optimisations were performed for each cluster size, and a range of both $d$ and $f$. When $d_+>0$ the probability of encountering the known $O_h$, non-icosahedral, ground state of the 38 atom cluster is very similar to that for the neighbouring 37 and 39 atom clusters, while for optimisations purely in normal space the $O_h$ 38 atom cluster is comparatively rarely encountered. It is notable that the increase in probability of encounter on exploiting hyperspace is particularly great for this supposedly difficult 38 atom cluster. Indeed, using GOSH it does not appear to be particularly challenging at all, and the probability of encountering the ground state is similar to that observed for minima hopping and evolutionary algorithms.\cite{schonborn2009performance}

Locating the ground state of the 75 atom Lennard-Jones cluster is certainly challenging, and the statistics are not of sufficient quality for inclusion in Table \ref{ljclusters-2}. However, using GOSH it is possible to repeatedly encounter it. For $d=3+1$ and $f=0.1$, based on 5 encounters, $-\log_{10}(p_e)=8$.

GOSH appears to be relatively insensitive to the details of the preparation of the initial random structures. The packing fraction, $f$, of the initial random structures is seen to strongly impact $p_e$ for conventional random search ($d_+=0$), but less so for $d_{+}=1$. In these tests increasing the number of extra dimensions from one to two does not significantly change $p_e$ for most of the cluster sizes tested (although it does appear to be somewhat diminished for $d=3+2$ and the 37 and 38 atom clusters).
\begin{figure}[]
\centering
\includegraphics[width=0.45\textwidth]{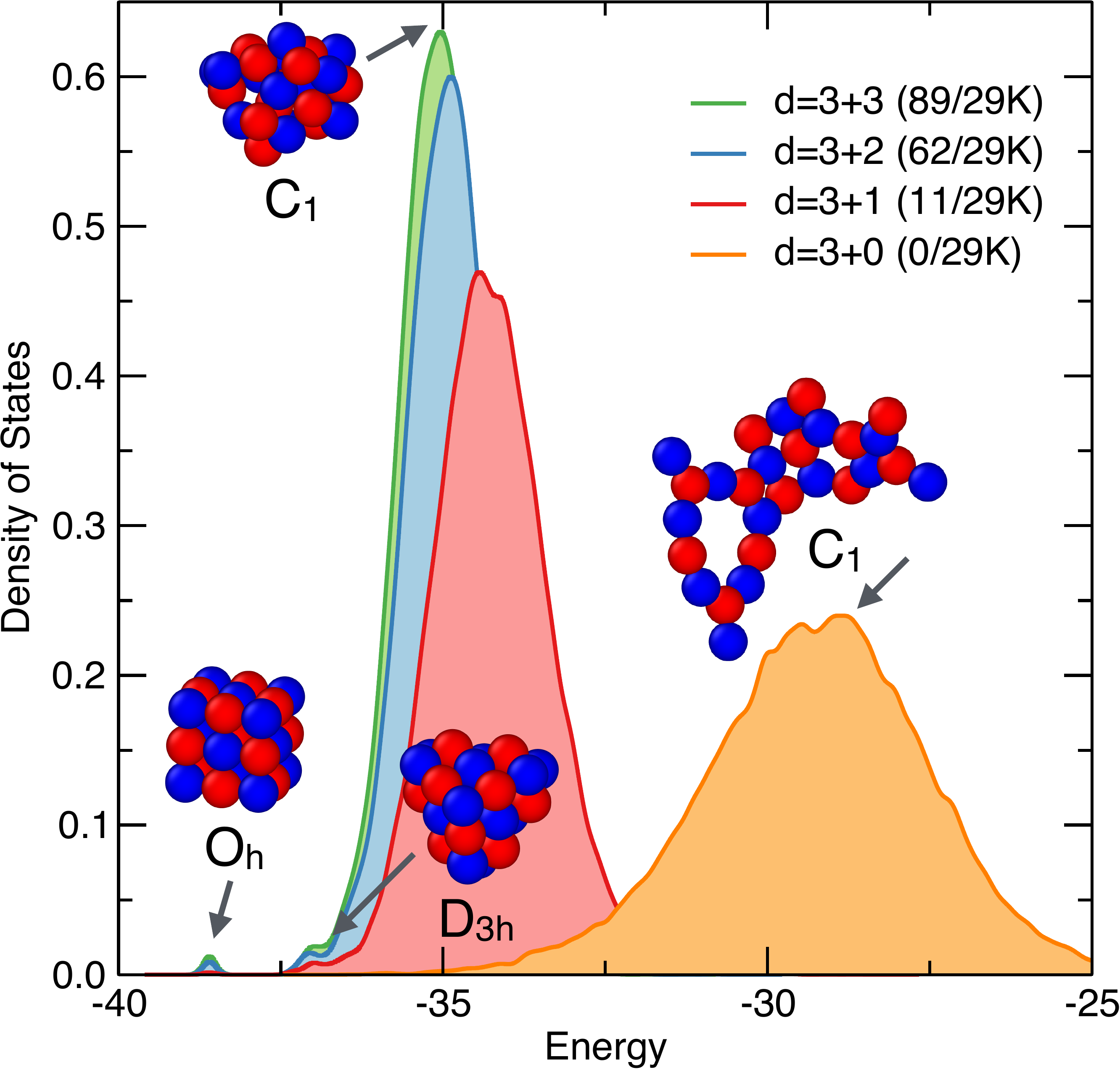}
\caption{Density of states for a model 3d binary system, with 13 atoms of type A (red), and 14 atoms of type B (blue) and a low packing fraction of $f=0.05$ for the initial random structures. }\label{cube}
\end{figure}

\subsection{Binary systems}
\label{bs}

The performance of GOSH and RSS for determining the low energy structures of multiple species is investigated for a system which is constructed so as to be extremely challenging for random search - the identification of the ground state for a binary cluster in one dimension, as might be found encapsulated within a nanotube.\cite{wynn2017phase} The cluster consists of atoms with equal size of type A (red, ``positive'') and B (blue, ``negative'') with $N_A=12$ and $N_B=13$.  When $d_{+}=0$ the probability of randomly encountering the alternating ground state depends on the initial structure having the correct ordering,  and is very small ($p_e=\frac{N_A!N_B!}{(N_A+N_B)!}\approx10^{-6.7}$) as there is no way for the atoms to reorder during the optimisation. As can be seen in Figure \ref{wire1}, when $d_{+}=2$, the atoms can move past each other to find the energetically favourable ordering, and $p_e\approx10^{-1.2}$ (see Figure \ref{wire2}). In Figure \ref{wire3} the performance of GOSH with system size is explored. For $d_+=0$ the probability of encountering the groundstate is low, and a high packing fraction ($f=0.75$ ) is required to approach the theoretical probability of correct ordering. Increasing $d_+$ to 1 (and then to 2) results in a dramatic acceleration. For $N_A=16$ and $d_+=2$ the groundstate is encountered twenty million times more frequently than possible for $d_+=0$. This acceleration is expected to become greater with the further increase of $N_A$.

\begin{figure}[]
\centering
\includegraphics[width=0.45\textwidth]{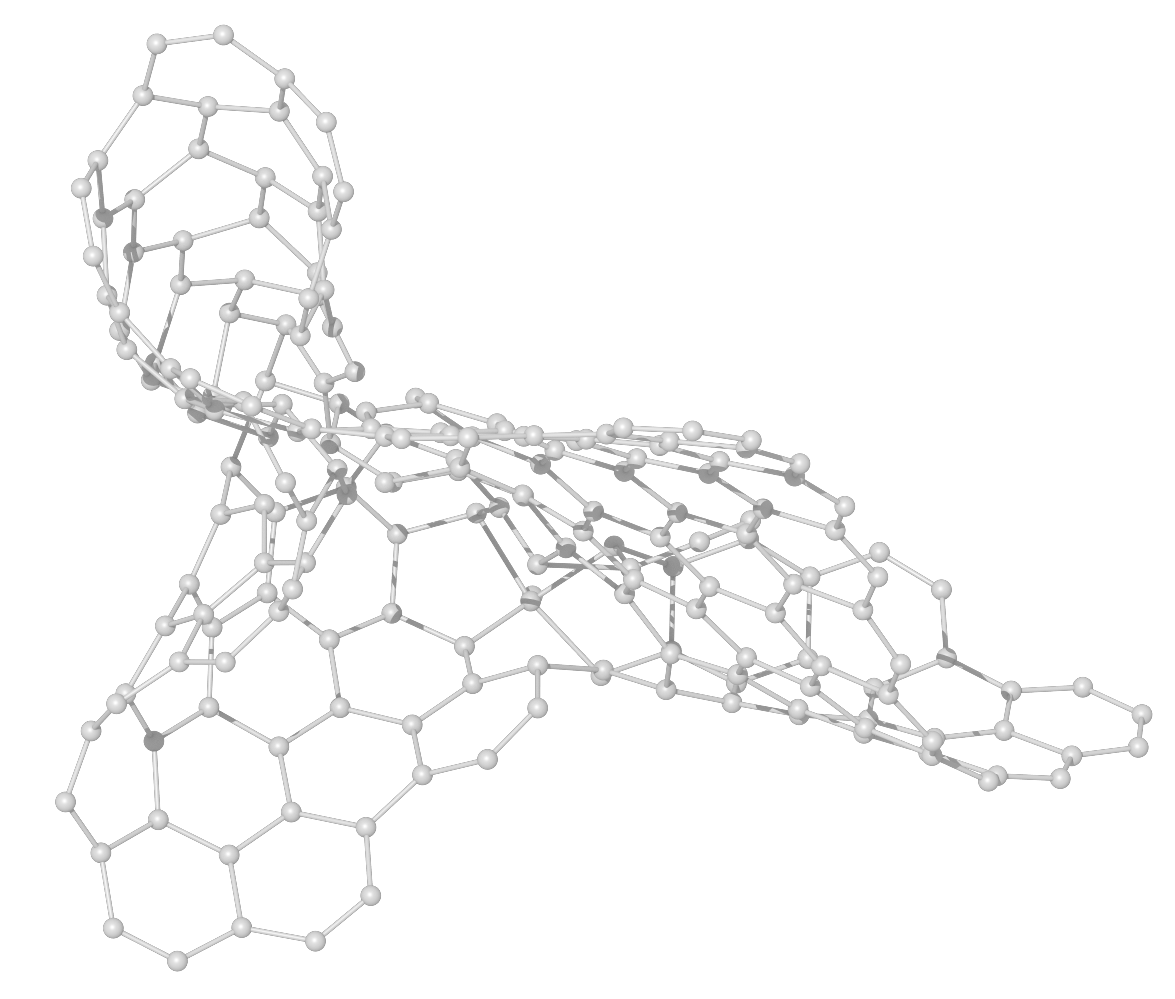}\\
\caption{Local minimum of a model system with defined connectivity, and $d_{+}=0$. Random initial configurations typically relax into topologically frustrated configurations.}\label{net-tangle}
\end{figure}

This one dimensional example is an extreme case, but a similar acceleration is found for a three dimensional binary cluster ($N_A=13$ and $N_B=14$) -- see Figure \ref{cube}. The global minima is a cubic $O_h$ symmetry cluster and a metastable $D_{3h}$ cluster is found at higher energy. When $f=0.05$ the maximum of the distribution moves to lower energy as $d_{+}$ is increased, with a dramatic increase in the probability of encountering the ground state as $d_{+}$ is increased from $0$ (where it is not encountered at all) to $1$. For $d_{+}=0$, a higher packing fraction of $f=0.3$ does increase the probability of locating the $O_h$ cluster to $14/29$K, but the $D_{3h}$ cluster is not found at all. This suggests that excessively exploiting packing to bias the search toward low energy structures prevents the  metastable $D_{3h}$ cluster from being located. 

\subsection{Connected systems}
\label{cs}

\begin{figure}[]
\centering
\includegraphics[width=0.45\textwidth]{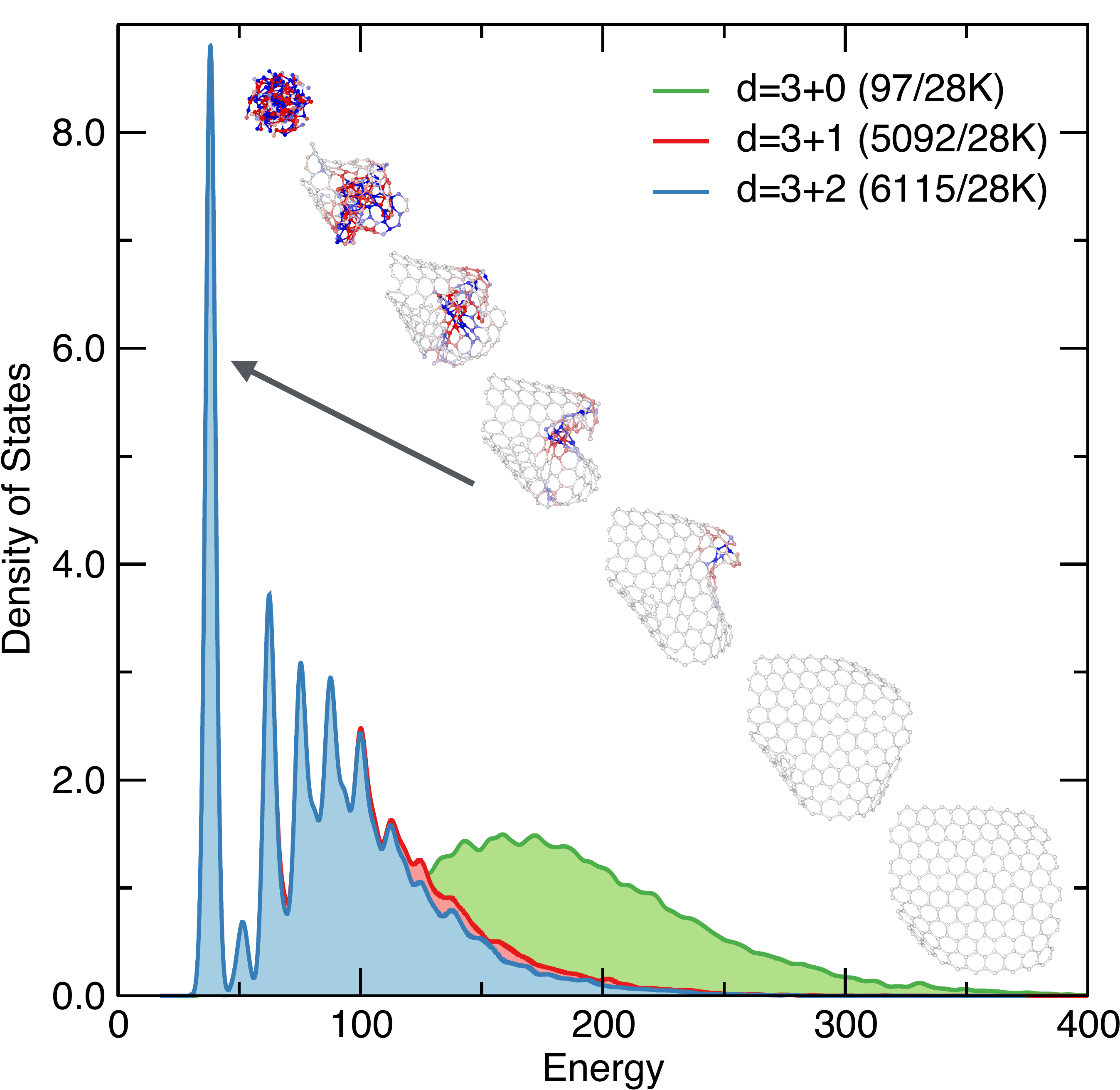}\\
\caption{Model networked system. The packing fraction of the initial random structures is $f=0.1$. A sequence of snapshots for an optimisation of a structure whose local minimum corresponds to a low energy sheet, for $d_{+}=1$, is presented. See Fig. \ref{net2} for further details.}\label{net}
\end{figure}

To explore the performance of GOSH and RSS for strongly covalently bonded systems 207 carbon atoms are cut from a graphene sheet to produce an approximately square mat. The connectivity in the mat is recorded in an adjacency matrix which is used to distinguish between the bonded and non-bonded interactions described by the interatomic potential detailed in Section \ref{iat}. Figure \ref{net-tangle} shows a typically topologically frustrated local minimum. In Figure \ref{net} it is shown that the probability of encountering the low energy sheet, with the correct predetermined connectivity and topology increases by a factor of sixty when $d_{+}$ is increased from $0$ to $1$. There is a weak dependence on increasing $d_{+}$ further to $2$.

\begin{figure}[]
\centering
\includegraphics[width=0.45\textwidth]{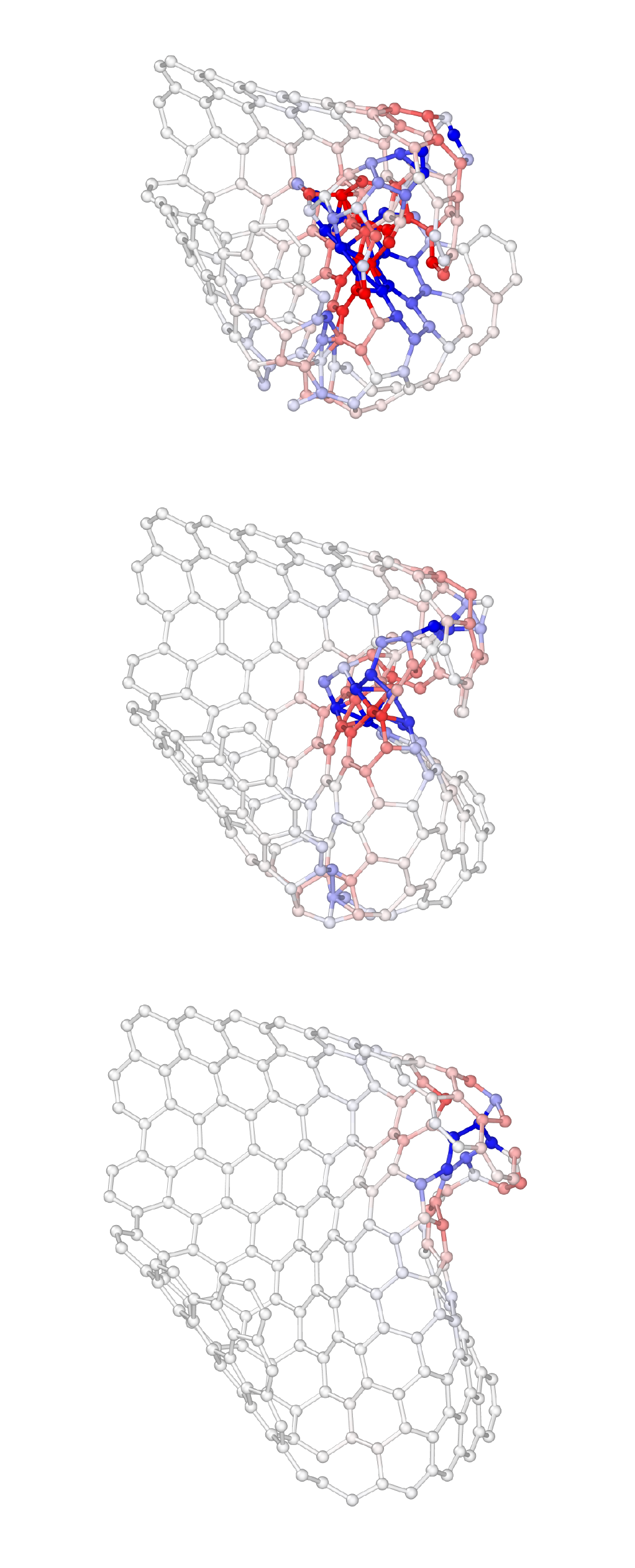}\\
\caption{Snapshots of a geometry optimisation of a model system with defined connectivity, and $d_{+}=1$. The atoms are coloured according to their position in the extra dimension - red for positive values, and blue for negative ones. The locally optimal atoms rapidly leave hyperspace. Where the network remains tangled the atoms continue to inhabit the extra dimension, and appear to move through each other in normal space. See Ref. \onlinecite{si} for an animation of the optimisation.}\label{net2}
\end{figure}

In Figure \ref{net2} an optimisation to a low energy configuration is followed in detail. Regions of the system for which the coordination is well satisfied rapidly collapse into three dimensional space. Where there is frustration, and the network passes through itself, or is knotted, the collapse is delayed, and the additional dimension is exploited in the optimisation to a lower energy configuration.

\section{Discussion}

It would appear that extending the search for low energy arrangements of atoms into hyperspace provides a system, on its descent from some energetic starting point in configuration space, a degree of ``farsightedness'' in the normal space. The deeper valleys and minima that lay beyond an obstruction in the normal space are ``felt'' and the direction of the descent adjusted accordingly. Entering into hyperspace, the system ``stands up'' and has a good look around for promising directions in which to head. At the same time, these excursions into hyperspace can help the avoidance of knots and other tangles. Mathematically, all knots in one dimensional objects (for example, a linear polymer chain) are trivial in four dimensions. Similarly for higher dimensional objects, such as knotted sheets (see Figure \ref{net-tangle}) in five dimensions.\cite{ranicki2013high}

The no-free-lunch theorem\cite{wolpert1995no} warns us against expecting too much from global optimisation strategies, and their general performance. However, the apparent efficiency of GOSH with RSS when applied to the specific task of locating the low energy arrangements of atoms is reached without excessive tuning,  and with few additional parameters, while preserving parallelism. Up until this point only the most minimal adjustments to the parameters $d_+$, $\mu_0$ and $\beta$ have been made in an attempt to give a balanced picture of the performance of GOSH when combined with RSS. It is, however, interesting to consider what the potential performance of GOSH might be. Indeed, changing $\mu_0$ and $\beta$ from 10 and 1.001 (for $d_+=1$ and $f=0.1$) to 15 and 1.0001 respectively decreases $-\log_{10}p_e$ from 3.3 to 2.8 for the encounter of the Lennard-Jones 38 atom $O_h$ cluster. This is lower than that found ($-\log_{10}p_e$=3.1) for both minima hopping and evolutionary algorithms in Ref. \onlinecite{schonborn2009performance}, and suggests the potentially excellent performance of GOSH.

GOSH could also be combined with learning algorithms if desired, or in combination with the constraints used in the generation of more ``sensible'' random structures (such as symmetry, fragments and distances). As shown in Figure 1, the relax and shake (RASH,\cite{pickard2011ab} a zero temperature basin hopping\cite{leary2000global} with a move consisting of an overlap avoiding random motion of all atoms with a chosen amplitude) can straightforwardly integrate GOSH. Using RASH ($N_{\rm max}$=1000, $r_{\rm amp}$=0.4, $f$=0.1, $d_+$=1, $\mu_0$=20 and $\beta$=1.001) the $-\log_{10}p_e$ for the ground state of the 75 atom Lennard-Jones cluster is reduced from 8 for RSS to 5.8, and high quality statistics can be collected. The probability of encountering the ground state 55 atom Lennard-Jones cluster on combining GOSH with RASH ($N_{\rm max}$=10, $r_{\rm amp}$=0.4, $f$=0.1, $d_+$=1, $\mu_0$=10 and $\beta$=1.001) is increased beyond those reported in Ref.\onlinecite{schonborn2009performance}, with $-\log_{10}p_e$=1.9.

Travels from hyperspace are not cost free. The number of steps taken will often be greater, as the system covers larger distances in configuration space, avoiding traps which would otherwise curtail the optimisation. The computational overhead for treating the extra dimensions is relatively minor if $d_+$ is not large. One might expect that increasing $d_+$ leads to increasing freedom, and a higher chance of encountering the low energy configurations. While this generally appears to be the case, there are exceptions (the 37 and 38 atom Lennard-Jones clusters in this study) and when taking into acount the increased computational cost, choosing $d_+=1$ appears to offer very good performance in the tests. It should be noted that benchmarking GOSH for different $d_+$ is challenging for systems that are sensitive to the initial packing, as for a fixed $f$ a larger $d_+$ implies a greater effective degree of packing, and a greater bias in the search. This might explain some of the enhanced performance seen with increasing $d_+$  from 1 to 2 in the binary $d_0=3$ cluster. However, the increased packing on entering hyperspace is achieved while maintaining a diversity (or randomness) that can be absent in highly packed structures in normal space. For example, for the 38 atom Lennard-Jones cluster a sufficiently high packing fraction ($f$=0.525) guarantees the identification of the O$_h$ ground state. But for 37 atoms the ground state is never found (for $f$=0.5125). In the 1D binary Lennard-Jones example, packing is essential to approach the theoretical maximum performance in normal space, but GOSH goes well beyond it, with low packing fractions.

In this work the additional dimensions have been treated on an equal footing. However, it would be interesting to explore the impact of more complex schedules for their removal. For example, $\beta$ could be chosen so as to be different for each of the extra dimensions, as could $\mu_0$. Or the extra dimensions might be removed sequentially, one after the other.

Existing codes are typically restricted to three or fewer dimensions, requiring a bespoke code to be written to perform the tests presented here. This code currently treats simple interatomic potentials, based on distances. But this could be extended to angles, which are well defined in higher dimensions through scalar products. It would appear straightforward to extend GOSH to more realistic model potentials, periodic boundary conditions and constraints such as symmetry. This will be the immediate focus of future work, which will allow GOSH to be applied to empirical models of systems varying from minerals to complex biological molecules.

The true power of structure prediction has been revealed through the marrying of stochastic search algorithms\cite{pickard2006high,oganov2006crystal,wang2012calypso} with modern first principles, plane wave DFT codes.\cite{clark2005first,kresse1996efficient} The modification of such complex codes to operate in hyperspace will be involved. An immediate route to obtain first principles accuracy and robustness, while remaining within the framework of interatomic potentials, may well be through machine learned models, which have recently been coupled with structure searching,\cite{ouyang2015global,eivari2017two,deringer2017extracting,deringer2018data} and their extension to hyperspace.  Although formulated here for the determination of low energy arrangements of atoms, it is possible that hyperspatial optimisation will prove to be useful in determining the optimal arrangements of more generally shaped objects in space, such as polyhedra and other nonspherical objects.\cite{torquato2009dense,damasceno2012predictive} The reduction of highly dimensioned data to two or three which might be readily be visualised is a task of increasing importance in materials informatics.\cite{ceriotti2011simplifying, isayev2015materials} The algorithms typically involve a global optimisation step for arrangements of points interacting through forcefields, and may well be accelerated by GOSH.

\section{Conclusion}

To conclude, structural optimisations starting in hyperspace can avoid traps in the energy landscape and accelerate structure prediction. The energy landscape is extended to additional spatial dimensions, and structure optimisations are performed on this landscape as an energy penalty for entry into the additional dimensions is increased. When the penalty is large enough, the locally optimal structures are to be found entirely in normal space, and tests show that the probability of reaching low energies is much enhanced. The approach maintains the parallelism of random structure searching, and may be combined with more complex schemes for structure prediction which depend on local structural optimisation.

\begin{acknowledgments}
CJP is supported by the Royal Society through a Royal Society Wolfson Research Merit award and the EPSRC through grants EP/P022596/1, and thanks Bartomeu Monserrat for his careful reading of the manuscript.
\end{acknowledgments}

%

\end{document}